\documentclass[12pt]{article}

\textheight by \topskip \textwidth 470pt

\makeatletter
\@addtoreset{equation}{section}
 
\makeatother

\usepackage{amsmath,amssymb}
\begin{document}


%
%

\title{Generalized commutation relations and Non linear momenta theories, a close
relationship }
\author{{H\'ector Calisto and Carlos Leiva}\thanks{
E-mail: hcalisto@uta.cl}\thanks{%
E-mail: cleivas@uta.cl} \\
{\small \textit{Departamento de F\'{\i}sica, Universidad de Tarapac\'{a},
Casilla 7-D Arica, Chile}}}
\maketitle

\begin{abstract}
A revision of generalized commutation relations is performed,
besides a description of Non linear momenta realization included
in some DSR theories. It is shown that these propositions are
closely related, specially we focus on Magueijo Smolin momenta and
Kempf et al. and L.N. Chang generalized commutators. Due to this,
a new algebra arises with its own features that is also analyzed.
\end{abstract}


%
%











\section{\protect\bigskip Introduction}

The standard model is the general framework of our present
microscopical theories and it has already been shown that is very
successful. Its validity has been confirmed by experiments ranging
on a very wide range of energies, from $eV$ to $TeV$ \cite{1}.
However, from a very recent perspective \cite{2}, it seems to be a
low energy effective theory, because there are strong difficulties
in introducing gravity in the microscopic formulation; on the
other hand there are no reasons why it should be the right theory
at very high energies and one or all its ingredients (special
relativity, quantum mechanics and cluster decomposition principle)
could fail at those energies. Indeed, nowadays there are some
evidences that Lorentz invariance could not be an exact symmetry
at high energies, as recent developments in quantum gravity
suggest \cite{3,4}.

Today, one of the principal challenges is to combine successfully
the gravitational interactions with Quantum Mechanics. There are
direct proposals to do this: M theory and loop quantum gravity are
good examples, however we can not say if they are the right
choices. Meanwhile the analysis of features that a right theory
should have and problems that arise in different efforts, can be
very fruitful. A fundamental idea that has been proposed in order
to achieve this, is the existence of a fundamental length. A well
developed list of arguments in order to introduce a minimal length
is presented in the Hossenfelder paper \cite{4a}, a shorter one is
the following:

1. String Theory naturally includes a minimal length scale because
its very success arises from the fact that interactions do no take
place at one point in space time but they occur on the world
sheet. Meanwhile, generalized uncertainty relations have aroused
and this is not a minor feature.

2. In Loop Quantum Gravity \cite{4} the area operator has a discrete
spectrum which gives rise to a smallest-distance structure.

3.Non-Commutative Geometries \cite{5} modify the algebra of the generators
of space-time translations such that position measurements fail to commute.
The commutator is now proportional to a matrix that has a dimension of $%
length^2$

4. A minimal length can be found from the Holographic Principle, which
states \cite{6} that the degrees of freedom of a spatial region are
determined through the boundary of the region and that the number of degrees
of freedom per Planck area is no greater than unity. This leads to a minimal
possible uncertainty in length measurements \cite{7}.

5. Several phenomenological examinations of possible precision measurements
\cite{8}, thought experiments about black holes \cite{9} or the general
structure of classical\cite{10}, semi classical \cite{11} and quantum-foamy
space-time \cite{7}. All of them lead to the conclusion that there exists a
fundamental limit to distance measurement.

A fundamental minimal length is closely associated with Non
commutativity, deformed Heisenberg algebra and DSR theories, all
of these theories are also related to Large Extra Dimensions
theories and a revision is really fundamental. A brief synopsis of
theses areas are presented in the next sections 2 and 3, and the
relationship between Non linear momenta proposed by Magueijo
Smolin and the Generalized Uncertainty Principle is shown in
section 4. Finally, a discussion is presented in section 5.


\section{A brief review on Non commutativity and Deformed commutation
relations}


The first example of a noncommutative space that was clearly
recognized in physics is the quantum phase space. In fact the
first considerations on their quantified differential geometry
were developed, already 1926, by P.A.M. Dirac \cite{12,13}. In
these works, Dirac discovered the algebraic structure of the
quantum phase space postulating its famous rule of quantization of
a classic theory, that consists in the replacement of the Poisson
brackets of two classical observables by $\mathrm{i}\hbar $ times
the commutator of the associated quantum operators. In this way
the coordinates of the phase space $p$ and $q$ are transformed
into
noncommutative operators $\hat{p}$ and $\hat{q}$ whose commutator is $%
\mathrm{i}\hbar$. This Non commutativity implies a relation of
incertitude between the operators $\hat{p}$ and $\hat{q}$, that
makes disappear the notion of individual points in the phase space
being \textit{the Bohr cell} the more plausible idea that
survives. In the limit $\hbar \rightarrow 0$ it recovers the
ordinary phase space.

This particular algebra of operators inspired later the more
radical idea which consists in the replacement of the coordinates
$x_{\mu } $ of the space time by noncommutative operators. As it
happens in the previous case, the relation $[x_{\mu }, x_{\nu
}]\neq 0$ implies an incertitude principle between the coordinates
of the space time that makes to disappear the image of point to
short distances. One can argue that as the Bohr cell replaces the
points of the classic phase space, the appropriate intuitive
notion to replace a point is \textit{the Planck cell} of
dimensions given by Planck's area.

Pauli in a revision of the basic principles of quantum mechanics
\cite{14} affirmed that only relativistic quantum mechanics is
logically complete and expressed vigorously his conviction that
new limitations in the possibilities of measurements  would have
to be expressed more directly in a future theory, and that these
would be associated with an essential and deep modification of the
basic concepts and the formalism of present quantum theory. Pauli
also affirmed that the concepts of space and time on very small
scales need a fundamental modification.

The origin of this is the fact that the calculated values of some
observables grows to infinity when the continuous theories are
extrapolated until arbitrarily small distances, although the
measured values are in fact finite. This appeared first in the
classical theory of electromagnetism.

Quantum electrodynamics attenuated these divergences, but did not
remove them. With the purpose of controlling these divergences,
around 1930, Heisenberg \cite{15,16}, proposed to replace the
continuous space time by a discrete structure. Nevertheless, at
first glance a discrete structure breaks the relativistic
invariance that, at that time, was a fundamental requirement of
any theory. Later Snyder \cite{17}, suggested the idea to use a
noncommutative structure, and showed that this necessarily implies
the existence of a length scale below which the notion of point
does not exist any more. The Snyder method maintains invariance
with respect to Lorentz transformations and it is possible that
introducing it in a field theory provides an effective cut-off,
that is to say, one minimal length in the space time to which the
theory is sensible, eliminating therefore the infinities. The
Snyder theory later was corrected by Yang \cite{18} to include
translational invariance and after some initial variations and
developments \cite{19,20,21,22,23,24} this idea practically was
forgotten, mainly because the renormalization program was revealed
appropriate to predict indeed finite numerical values for the
observable magnitudes in quantum electrodynamics, without
resorting to the non commutativity.

John Von Neumann was the first in trying to describe such quantum spaces
rigorously and introduced the term noncommutative geometry to talk about to
a geometry in which an algebra of functions is replaced by a noncommutative
algebra. The idea of noncommutative geometry were retaken in the Eighties by
the French mathematician Alain Connes \cite{25}, that generalized the notion
of differential structure to the noncommutative case, that is to say, to
arbitrary algebras. Connes defined a generalized integration, this took to
the description of the noncommutative space time and allowed the definition
of theories of field in such spaces.

A different approach has been suggested since long ago, and this
idea to generalize Quantum Mechanics via commutation relation has
 originated from the following arguments:

\begin{itemize}
\item \underline{Esthetics}: There is an obvious, unsatisfying asymmetry
between the current treatment of the macro and the micro world, geometry
being given a priori at the sub nuclear level. Geometry is assumed to be
independent of the physical phenomena.

\item \underline{Curiosity}: Can commutation relations be the same at
energies characteristic of atomic at present day physics, that is, over the
range of some twelve order of magnitude?.

\item \underline{Phenomenology}: Might the striking high energy phenomena
such as quark confinement and the regularities exhibited by the "heavy
photons" be connected with the geometry of space?.
\end{itemize}

Then to do research on the above mentioned questions \cite{32} the following
generalization of the commutation relation was proposed:
\begin{equation}
\left[ q,p\right] =\mathrm{i}\hbar +\frac{\mathrm{i}\ell }{c}\mathcal{F}(q,p)
\end{equation}
where $\ell$ is a constant with dimensions of length which is
sufficiently small so that the second term in the right hand side
is negligible except for very high energy or momentum, in other
words, for low energies the usual Heisenberg commutation relation
holds, and $c$ is the velocity of light in vacuum.

We assume that the variable $p$ well understood "the momentum
measured in the laboratory" and therefore we consider equation (1)
as an equation which determines the position operator $q$ when
$\mathcal{F}(q,p)$ is given. In the low energy or low momentum
limit $q=\mathrm{i}\hbar d/dp$ is the usual position operator.

The function $\mathcal{F}(q,p)$ should in general depend on the dynamics of
the problem, and therefore the operator $q$ and the $q-$eigenvalues, this is
to say, the physical space, is not given a \underline{priori} but is
determined by the physics of the problem as represented by the choice of $%
\mathcal{F}(q,p)$. Hence the name to describe this procedure: \underline{%
Dynamical Quantization}.

We further assume the existence of a Hamiltonian function $H=H(q,p)$, and
the validity of the Heisenberg equation of motion:
\begin{equation}
\frac{d\Omega}{dt}=\frac{\mathrm{i}}{\hbar}\left[H,\Omega\right]
\end{equation}
where $\Omega$ is a dynamical variable and $H$ is the Hamiltonian of the
system, which continues being valid.

Next we will briefly survey some previous results, with some of
the possible elections of $\mathcal{F}(q,p)$ both, for the
non-relativistic case and a
possible relativistic generalization. The simplest choice for $\mathcal{F}%
(q,p)$ is $H(q,p)$ which leads to the commutation relation:
\begin{equation}
\left[ q,p\right] =\mathrm{i}\hbar +\frac{\mathrm{i}\ell }{c}H(q,p)
\end{equation}%
As was postulated in \cite{26,27}. Equation (2.3) implies
\begin{equation}
\Delta p\Delta q\geq \left\vert \frac{\hbar }{2}+\frac{\ell }{2\,c}%
E\right\vert
\end{equation}%
then the lower bound of $\Delta p\Delta q$ grows linearly with
energy, except for logarithmical corrections as it was confirmed
by Giffon an Predazzi \cite{33} using high energy data, which
preliminarily yielded $\ell =2.3\times 10^{-6}$fm. The harmonic
oscillator problem can still be solved exactly with equation
(2.3), yielding a non-equal spacing law for the energy levels
\cite{27}.

\begin{equation}
E_{n}=\frac{\hbar \omega }{2\alpha }\left[\frac{2-\left( \frac{\beta }{%
\alpha }\right) ^{n}-\left( \frac{\beta }{\alpha }\right) ^{n+1}}{1-\frac{%
\beta }{\alpha }}\right]
\end{equation}%
where $\alpha =1-\omega \ell /2c$ and $\beta =1+\omega \ell /2c$.

In the one dimensional case, for a non relativistic free particle we have
\begin{equation}
[q,p]=\mathrm{i}\hbar(1+\delta^2p^2)
\end{equation}
where $\delta^2=\ell/2\,m\,\hbar\,c$. The corresponding position operator
eigenvalues problem leads to the physical space. The result was that the
spectrum is discrete \cite{28}
\begin{equation}
\lambda_n=2\,n\,\hbar\,\delta,\qquad n=0,\pm 1,\pm 2,\cdots
\end{equation}
That is to say, the space generated by $H=p^2/2\,m$, is a one-dimensional
lattice, the smallest interval being
\begin{equation}
\Delta q_{min}=2\hbar\delta=\sqrt{\frac{2\,\hbar\,\ell}{m\,c}}
\end{equation}
Then,  once the particle has been located at any arbitrary point
of the one dimensional space, the rest of the space "feels" it,
the lattice appears instantaneously; in this sense, the geometry
acts as constant force, a linear potential. Further, if $\Delta
q_{min}$ is the spatial extension of an extended object, then it
is meaningless to ask for its "constituents" , that is, objects of
smaller size, because no test particle can go into it.

For a relativistic generalization our starting point is the observation that
the function $\mathcal{F}$ takes place in (6) and can be written as follows
\begin{equation}
\mathcal{F}=\frac{p^2}{2\,m}=\frac{1}{2\,m}\,p\cdot p
\end{equation}
which suggests to choose, for the relativistic free particle case
\begin{equation}
\mathcal{F}=\frac{p_\mu p_\nu}{2\,m}
\end{equation}
Equations (1) and (10) then lead to the following generalization
\begin{equation}
\left[q_\mu,p_\nu\right]=-\mathrm{i}\hbar\left(\,g_{\mu\,\nu}- \delta^2{%
p_\mu\,p_\nu}\right)
\end{equation}
where $g_{0\,0}=-g_{k\,k}=1$ for $k=1,2,3$ and $g_{\mu\,\nu}=0$ para $%
\mu\neq\nu$. The implication of this commutation relation can be
investigated. Assuming as before that the $p_\mu$ are given we find that
\begin{equation}
{q}_{\mu }=-\mathrm{i}\hbar \left( \,g_{\mu \,\nu }-\delta ^{2}{p_{\mu
}\,p_{\nu }}\right) \frac{\partial }{\partial p_{\nu }}+\mathrm{i}\hbar
\,\kappa \,p_{\mu }  \label{12}
\end{equation}
where $\kappa$ is a real positive constant which determines the
weight function in the inner product if we wish to give a physical
meaning to our operators. By imposing
$(q_\mu\phi,\psi)=(\phi,q_\mu\psi)$ we find that the suitable
internal product is
\begin{equation}
\left( \psi ,\phi \right) =\int \,d\tau \,\frac{\psi ^{\ast }\,\phi }{\left(
1-\delta ^{2}\,g_{\mu \,\nu }p_{\mu }\,p_{\nu }\right) ^{1-\gamma}}
\end{equation}
where $d\tau$ is an appropriate volume element, $\gamma =\frac{\kappa }{%
\delta ^{2}}-\frac{1}{2}\,(D-1)$ and $D$ is the dimension of space time. We
remark that since $\kappa$ is a free parameter we can choose $\beta$ so that
the internal product normally used in quantum mechanics continues being
valid.

The main results of this generalization are:

\begin{itemize}
\item Coordinates operators do not commute, their commutators being
proportional to the infinitesimal generators of the Lorentz group.

\item The spatial operators $q_1$, $q_2$, $q_3$ having discrete spectrum.

\item The time operator $q_0$ has a continuum spectrum.
\end{itemize}

Applications of this type of generalization, show that the energy spectrum
of the one dimensional harmonic oscillator is given by
\begin{eqnarray}
E_{n}=\hbar \omega \left[ \left( n+\frac{1}{2}\right) \,\sqrt{1+\left( \frac{%
\omega \ell }{2\,c}\right) ^{2}}\right] +\hbar \omega \left[ \left( n^{2}+n+%
\frac{1}{2}\right) \frac{\omega \ell }{2\,c}\right]
\end{eqnarray}
and the eigenfunctions are the Gegenbauer Polynomials. In these calculations
we have used the one dimensional version of the position operator \ref{12}.

The eigenvalues problem for the three-dimensional harmonic
isotropic oscillator can be solved using the three-dimensional
version of position operator \ref{12}. We found
\begin{eqnarray}
E_n =\hbar\omega\left(n\!+\!\frac{3}{2}\right) \sqrt{1\!+\!\left(\frac{%
\omega\ell}{2\,c}\right)^2}\ + \hbar\omega\left(n^2\!+\!3\,n-s(s+1)\!+\!%
\frac{3}{2}\right)\frac{\omega\ell}{2\,c}
\end{eqnarray}
where $s(s+1)$ are the usual eigenvalues for the angular orbital
momentum of the oscillator. The eigenfunctions are the Jacobi
Polynomials.

Following suggestions originated in quantum theory of gravity and
string theory, Kempf, Mangano and Mann \cite{33} have suggested a
generalized uncertainty relation of the form
\begin{equation}
\left[ \mathbf{x},\mathbf{p}\right] =-\mathrm{i}\hbar \left( \,\mathbf{1}%
+\alpha \,\mathbf{x}^{2}+\beta \,\mathbf{p}^{2}\right)
\end{equation}%
where $\alpha $, $\beta $ are positive and independent parameters.
They discussed some consequences of this generalization in non
relativistic quantum mechanics and have worked in detail the case
$\alpha =0$. Their result are very similar to those found in
\cite{19}. Following this same type of suggestions and with
similar results, Chang and coworkers also have developed some
applications to non relativistic quantum mechanics \cite{34,35}
and they have discussed some consequences of the non commutativity
in classical mechanics \cite{36}. Preliminary applications to
relativistic quantum mechanics and quantum fields theory have been
developed recently by Yamaguchi in a Riemannian energy-momentum
space \cite{37,38} and by Ahluwalia-Khalilova\cite{DVA}.

The non commutative theories also play an important role in the
area of condensed matter, which is not only the concrete
accomplishment of the mathematical models used to explore the
properties of space time in physics at high energies and quantum
theory of gravity, but  represents concrete applications in an
area of increasing interest and impact. A classical example is the
theory of an electron in an external magnetic field, projected on
the lower Landau level, which can be treated like a non
commutative theory. It is by this reason that these ideas are
relevant for the study of the quantum Hall effect \cite{39} and in
fact they have been very useful in this context \cite{40}.

A recent and very convincing example of a noncommutative theory in the area
of condensed matter is the quantum theory of mesoscopic electrical circuits
developed by Li and Chen \cite{41,42} that takes into account the
discretization of the electric charge which leads to a new commutation
relation between the charge and current operators, that is similar to the
one studied in physics at high energies and quantum theory of gravity.
Several advances and applications in the context of the mesoscopic circuits
with discrete charge have been made by J.C. Flores \cite{43} and by J.C.
Flores in collaboration with C. Utreras \cite{44}.

Summarizing, noncommutative theories have been revealed as tools of
certain utility in theoretical physics. They appear as much in the
physics of high energies, for the description from a fundamental
level of the space time on small scale, as also in the area of
condensed matter to describe  the Hall effect and in the quantum
theory of the mesoscopic electrical circuits. The enormous activity
around these theories mainly is bound to the appearance of non
commutativity in the limit of low energies of string theory
mentioned previously. Since string theory is the only well-known
theory that it could unify all the fundamental interactions, it is
possible that the problems of control of divergences in the quantum
theory of field and quantization of gravity are basically related by
means of some sort of non commutative.

\section{DSR theories}

Absolute values of length, time or energy are not, at first
glance,  in agreement with the Lorentz transformations, this point
has produced the idea of modifying Lorentz symmetry.

Several solutions to the problem on how to modify the Lorentz
boosts, have been proposed. In particular, a very interesting
solution was given  by Doubly (or Deformed), Special relativity
(DSR) theories \cite{45,46,47}. These theories are based on a
generalization of Lorentz transformations through the more broad
point of view of conformal transformations, they have two observed
independent scales: velocity of light and Planck length.

Usually a modification of Lorentz boosts in momentum space is performed,
however when this is done, retrieving the position space dynamics can be a
very hard task, due to the loss of linearity. Kimberly, Magueijo and
Medeiros \cite{48}, have proposed some methods to undertake this problem by
using a free field theory. So, this is a worth research aspect of DSR
theories, that is not well understood until today.

Another approach can be seen in \cite{49}, on the other hand, the
approach of Deriglazov \cite{50}is very interesting because he
starts from a conformal group, but the idea is different from the
one proposed here, because it is based in  position space and the
problem of retrieving the position space dynamics is not present
in that work.

Some approaches have been performed in order to identify AdS spaces as
arenas for DSR theories and the approach of this paper could shed some light
on that problem too.

Even though  DSR theories were of increasing interest because they
could be useful as a new
tools in gravity theories, in Cosmology as an alternative to inflation \cite%
{51,52}, and in other fields like propagation of light \cite{53},
that is related, for instance, to cosmic microwave background
radiation, nowadays they fell out of favor with researches because
there ar serious conceptual issues that DSR has so far failed to
address like:

\begin{itemize}

\item They seem to belong to a trivial k deformations of
Poincar\'{e} algebra \cite{KGV}

\item Special relativity  extensions do not need to have  a non linear
character. \cite{DVA2}

\item They have very hard problems with multi particle sector (the
well known soccer ball problem) \cite {RW}

\item A fundamental length scale does not violate Lorentz
invariance as Snyder himself shown first \cite{17} and can be seen
also in \cite{Chryss}

\item Lorentz violations due to the combination of some
consequences of DSR like theories and elementary particle
interactions are at the percent level  some 20 orders of magnitude
higher than expected unless a very unnatural tuning is performed
\cite{CP}

\end{itemize}

Despite  all these problems, we will show that a conspicuous
feature of these theories is closely related to Generalized
Commutation Relations, the Magueijo Smolin non linear momenta that
are still fashionable.

 About this feature, Leiva \cite{54,55,56} has shown that it is
possible from a formal point of view to obtain Fock-Lorentz
\cite{57,58} and Magueijo-Smolin deformations, through a reduction
process and using conformal group generators as the generators of
the deformed Lorentz algebra. Then, the open problem of obtaining
the space time dynamics is solved through the relationship, on the
physical surface, between momenta and velocities that rise up from
this method.

More specifically, it was conjectured that the deformations of the Lorentz
algebra performed in the position space (the Fock-Lorentz formulation), can
be treated as a transformation made by a linear combination of conformal
group generators. On the other hand, the momentum case (the Magueijo Smolin
formulation), can be understood as an analog process, but the inclusion of a
new generator is needed. This new generator can be obtained from the same
theory, and it completes the set of symmetries of the massless Klein Gordon
equation. Then, a DSR massless particle is shown to be isomorphic to a
normal Lorentz particle living in a $d+2$ space, and the deformations are
induced by dimensional reduction.

\section{DSR and Generalized Commutation Relations}

It is possible to show a relationship between non commutativity,
deformed algebra and DSR theories. We are going to show that a
deformed algebra in the sense of \cite{45,46} and others, can be
seen as a first order approximation to some DSR theories. In fact,
if the commutator between the position operator and the
Magueijo-Smolin momentum operator $\pi$ is calculated, we obtain:
\begin{eqnarray}
[x_i,p_j]=i\delta_{ij},\quad i,j=1,\ldots n  \notag \\
\pi_j=\frac{p_j}{(1-l_{p}{p}^2)}, \ \ \quad\qquad  \notag
\end{eqnarray}

\begin{equation}
\lbrack x_{i},\pi _{j}]=i\frac{\delta _{ij}}{(1-l_{p}{p}^{2})}+i\frac{%
2l_{p}p_{i}p_{j}}{(1-l_{p}{p}^{2})^{2}}.  \label{18}
\end{equation}
and using the definition of $\pi _{j}$:

\begin{equation}
\lbrack \pi_{i},\pi _{j}]=0  \label{alg1}
\end{equation}
\begin{equation}
\lbrack x_{i},\pi _{j}]=2 i \delta _{ij}l_{p}\,f+2il_{p}\pi _{i}\pi _{j}
\label{alg2}
\end{equation}
where
\begin{equation}
f=\frac{{\pi}^{2}}{\sqrt{1+4l_{p}{\pi}^{2}}-1} \label{efe}
\end{equation}
at first order gives:
\begin{equation}
[x_i,\pi_j]\approx i\delta_{ij}(1+\ell_p {\pi}^2)+2i\ell_p \pi_i \pi_j
\end{equation}
and that  exactly is the relation proposed by Kempf et al. and
L.N. Chang.

On the other hand, using a very similar treatment to the one performed by
Leiva to obtain the DSR generators, J. Romero and A. Zamora \cite{59}
obtained the Snyder commutation relations. This is a new clue in that in all
these theories there is a very important underlying relationship.

The full position operator that satisfies \ref{18}, in the $\pi -$%
representation, can be easily calculated and we find
\begin{equation}
x_{i}=2\,i\,\ell _{p}\,f\,\frac{\partial }{\partial \pi _{i}}%
+2\,i\,l_{p}\,\pi _{i}\left( \pi _{j}\,\frac{\partial }{\partial \pi _{j}}%
\right) +2\,i\,\ell _{p}\,\kappa \,\pi _{i}
\end{equation}%
where $\kappa $ is a free parameter that can be fixed in the
definition of the internal product since as it can be easily
verified the operators $x_{i}$ given in (21) are not hermitian
with the internal product usually employed in quantum mechanics:
\begin{equation}
\left( \psi ,\phi \right) =\int \,d\pi _{1}\ldots d\pi _{n}\,\psi ^{\ast
}\,\phi
\end{equation}%
consequently, we needed to build an internal product in which the operators $%
x_{i}$ are self adjoint. We postulate the general form:
\begin{equation}
\left( \psi ,\phi \right) =\int \,d\pi _{1}\ldots d\pi _{n}\,\,\frac{\psi
^{\ast }\,\phi }{\mathrm{W}\left( \pi \cdot \pi \right) }
\end{equation}%
where $\mathrm{W}\left( \pi \cdot \pi \right) $ is a weight function
to be determined and $\pi \cdot \pi =\pi _{i}\pi _{i}$. Imposing the
condition of hermiticity of the operator $x_{i}$ with this new
internal product:
\begin{equation}
\left( x_{i}\psi ,\phi \right) =\left( \psi ,x_{i}\phi \right)
\label{24}
\end{equation}%
and requiring that the functions $\psi $ and $\phi $ vanish
suitably in the infinite, after an integration by parts, we find
that  \ref{24} is fulfilled, if the weight function $\mathrm{W}$
satisfies the following differential equation:
\begin{eqnarray}
f\,\frac{\partial W}{\partial \pi _{i}}+\pi _{i}\left( \pi _{j}\frac{%
\partial W}{\partial \pi _{j}}\right)  \notag \\
+ \left( -\frac{\partial f}{\partial \pi _{i}}+2\,\kappa \pi _{i}-(n+1)\pi
_{i}\right) \mathrm{W}=0
\end{eqnarray}
At this point and to simplify the equation for $\mathrm{W}$ we choose $%
\kappa =(n+1)/2$. Now using the function $f$ gives in \ref{efe} we
can see that:
\begin{equation}
\frac{\partial f}{\partial \pi _{i}}=\frac{2\pi _{i}\left( 1+2\ell _{p}\pi
^{2}-\sqrt{1+4\ell _{p}\pi ^{2}}\right) }{\left( \sqrt{1+4\ell _{p}\pi ^{2}}%
-1\right) ^{2}\sqrt{1+4\ell _{p}\pi ^{2}}}=2\pi_i\,\frac{\partial f}{%
\partial\pi}
\end{equation}%
Then it is evident that we can choose $\mathrm{W}$ as a function of $\pi =%
\sqrt{\pi_1^{2}+\cdots +\pi _{n}^{2}}$. With this ${\partial W}/{\partial
\pi _{i}}=\left( \pi _{i}/\pi \right) \partial \,\mathrm{W}/\partial \,\pi $
and the equation for $\mathrm{W}$ in spherical coordinates becomes
\begin{eqnarray}
\left( \frac{\sqrt{1+4\ell _{p}\pi ^{2}}}{\sqrt{1+4\ell _{p}\pi ^{2}}-1}%
\right) \pi \frac{d\,\mathrm{W}}{d\,\pi }  \notag \\
-\frac{2\left( 1+2\ell _{p}\pi ^{2}-\sqrt{1+4\ell _{p}\pi ^{2}}\right) }{%
\left( \sqrt{1+4\ell _{p}\pi ^{2}}-1\right) ^{2}\sqrt{1+4\ell _{p}\pi ^{2}}}%
\mathrm{W}=0
\end{eqnarray}%
this last equation can be integrated directly giving
\begin{equation}
W(\pi )=\frac{\sqrt{1+4\ell _{p}\pi ^{2}}}{1+\sqrt{1+4\ell _{p}\pi ^{2}}}
\end{equation}%
where a multiplicative integration constant has been chosen equal to one
because any constant of this type always can be incorporated in the
normalization constants of the wave functions $\psi $, $\phi $.

In order to complete this section we define the angular momentum operators
in the usual way as:
\begin{equation}
L_j=\epsilon_{j\,k\,\ell}\,x_k\pi_\ell
\end{equation}
where $\epsilon_{j\, k\, \ell}$ is the usual Levi-Civita symbol. Doing
simple algebraic manipulations we find that $L_j$ is explicitly given by:
\begin{equation}
L_j=-2\,i\,\ell_p\,f\,\epsilon_{j\,k\,\ell}\,\pi_k\,\frac{\partial}{\partial
\pi_\ell}
\end{equation}
In addition, other direct calculations allow to show that:
\begin{equation}
\left[x_j,x_k\right]=2\mathrm{i}\,\ell_p\left[1+2\left(1+\frac{\pi^2}{f}%
\right)\,\frac{\partial f}{\partial\pi}\right] \epsilon_{j\,k\,\ell}\,L_{%
\ell}  \label{alg3}
\end{equation}
The position operators do not commute. Evidently:
\begin{equation}
\Delta x_j\,\Delta x_k \neq 0
\end{equation}
as has been usual in this type of theories.

In this way equations (\ref{alg1}),(\ref{alg2}),(\ref{alg3}) are
the commutators of a new algebra that comprises Non linear momenta
in the way proposed by Magueijo Smolin and GUP theories.

\section{Discussion and outlook}

In section $2$ we have reviewed the principal arguments and ways
to introduce extra terms in the canonical relations. In section
$3$, we have seen some  facts  of DSR theories and in section $4$
we have seen that  how Magueijo Smolin momenta produce non linear
commutation relations that are, to first order, the same of those
proposed by Kempf et al. and Chang, used also by
Ahluwalia-Khalilova in an relativistic extension. Furthermore we
constructed explicitly the position operator and investigated its
features.

We think it is very important to study features that arise when a
minimal length or a Generalized Uncertainty Principle or Non
commutativity are introduced in areas like Classical Mechanics and
Quantum Mechanics, because new hidden or exotic symmetries could
emerge. \cite{Hor}

Finally,to extend the study to quantum field formulation could be
also a interesting task, indeed Lorentz violation in
supersymmetric Field Theories was  investigated by Nibbelink and
Pospelov \cite{60}. The extension of these theories to Quantum
Field Theory has been done just for the noncommutative case
\cite{61,62}, but they have not been deeply investigated, there is
just an extension  proposed by Magueijo \cite{63}.

\vskip 0.5cm \textbf{Acknowledgements} \vskip 5mm It is a pleasure for us to
thank Mauro Bologna and Fernando Mora by its valuable suggestions and
critics. Also we thank to the Research Grant 4723-05 from UTA which
partially supported this research.


\begin{thebibliography}{99}

\bibitem{1} J.M. Carmona and J.L. Cort\'{e}s, \textit{Infrared and
ultraviolet cutoffs of quantum field theory}, Phys. Rev. D 65 (2002) 025006
[arXiv:hep-th/0012028]

\bibitem{2} S. Weinberg. \textit{The quantum field theory of fields},
Cambridge University Press, Cambridge 199.

\bibitem{3} G. Amelino-Camelia, J. Ellis, N.E. Mavrotamos, D.V. Naonopoulos
and S. Sarkar,\textit{Test of quantum gravity from observations of $\gamma$%
-ray bursts}, Nature 393 (1998) 763.

\bibitem{4} C. Rovelli, Living Rev. Rel. 1,1 (1998); A. Perez, Class. Quant.
Grav. 20. R43 (2003); T. Thiemann, Lect. Notes Phys. 631, 41 (2003); A
Ashtekar and J. Lewandowski, Class, Quant. Grav. 21, R53 (2004).

\bibitem{4a} Hossenfelder, Sabine [arXiv:hep-ph/0410122]

\bibitem{5} R.J. Szabo, Phys. Rept. 378, 207 (2003).

\bibitem{6} G. 'tHooft, [arXiv:gr.qc/9310026], L. Susskind, J. Math. Phys.
36, 6377 (1995).

\bibitem{7} Y. J. Ng and H. van Dam, [arXiv:gr-qc/0403057].

\bibitem{8} E. P. Wigner, Rev. Mod. Phys. 29,255 (1957), H. Salecker and E.
P. Wigner, Phys. Rev. 109,571 (1958), J. D. Barrow, Phys. Rev. D 54, 6563
(1996), N. Sasakura, Prog. Theor. Phys. 102, 169 (1999).

\bibitem{9} F. Scardigli, Phys. Lett. B 452, 39 (1999); T. Padmanabhan, T.
R. Seshadri and T. P. Singh, Int. J. Mod. Phys. A 1, 491 (1986).

\bibitem{10} X. Calmet, M. Graesser and S. D. H. Hsu,
[arXiv:hep-th/0405033]; V. Daftardar and N. Dadhich, Int. J. Mod. Phys. A 1,
731 (1986); Y. J. Ng and H. van Dam, Mod Phys. Lett. A 9, 335 (1994).

\bibitem{11} -C. I. Kuo and L. H. Ford, Phys. Rev. D 47, 4510 (1993).

\bibitem{12} P. A. M. Dirac, \textit{The fundamental equations of quantum
mechanics}, Proc. Roy. Soc. \textbf{A109}, 642 (1926).

\bibitem{13} P. A. M. Dirac, \textit{On quantum algebras}, Proc. Camb. Phil.
Soc. \textbf{23}, 412 (1926).

\bibitem{14} W. Pauli: \textit{Die allgemeinen Prinzipien der Wellenmechanik}
in Handb.d.Physik (Springer, Heidelberg, 1933), 24/1, p. 246, 247, 271, 272.

\bibitem{15} W. Heisenberg, \textit{Die Leobachtbaren Grossen in der Theorie
der Elemntarteilchen}, Z Phys \textbf{120}, 513 (1943)

\bibitem{16} W. Heisenberg, \textit{Quantum Theory of Fields and Elementary
Particles}, Rev Mod Phys,\textbf{29}, 269 (1957)

\bibitem{17} H. S. Snyder, \textit{Quantized Space-Time}, Phys. Rev. \textbf{%
71}, 38 (1947).

\bibitem{18} C. N. Yang, \textit{On Quantized Space-Time}, Phys. Rev.
\textbf{72}, 874 (1947).

\bibitem{19} H. S. Snyder, \textit{The Electromagnetic Field in Space-Time},
Phys. Rev. \textbf{72}, 68 (1947).

\bibitem{20} A. Schild, \textit{Discrete Space-Time and Integral Lorentz
Transformations}, Phys. Rev. 73, 414-415 (1948). The Quantization of Space
and Time H. T. Flint Phys. Rev. 74, 209-210 (1948)

\bibitem{21} R. J. Finkelstein, \textit{On the Quantization of a Unitary
Field Theory}, Phys. Rev. 75, 1079-1087 (1949)

\bibitem{22} E. J. Hellund and K. Tanaka, \textit{Quantized Space-Time},
Phys. Rev. 94, 192-195 (1954)

\bibitem{23} E. L. Hill, \textit{Relativistic Theory of Discrete Momentum
Space and Discrete Space-Time}, Phys. Rev. 100, 1780-1783 (1955)

\bibitem{24} M. R. Hamilton and G. Sandri, \textit{Class of Fields in Snyder
Spaces}, Phys. Rev. D 8, 1788-1795 (1973)

\bibitem{25} A. Connes, \textit{Gravity coupled with matter and the
foundation of non-conmutative geometry}, (1996); hep-th/9603053.

\bibitem{26} I. Saavedra, \textit{A generalization of Quantum Mechanics for
High Energies and Quarks Physics}, in Quantum Theory and the Structures of
Time and Space, Vol. IV, Edited by L. Castell, M. Drieschner, and C. von
Weizsacker, Max Planck Intitut, Carl Hauser Verlag, Munich (1981).

\bibitem{27} I. Saavedra and C. Utreras, \textit{A generalization of Quantum
Mechanics for High Energies and Quarks Physics}, Phys. Lett., 98 B, 74
(1981).

\bibitem{28} S. Montecinos, I. Saavedra and O. Kunstmann, \textit{High
Energy Quantum Mechanics and Dynamical Quantization}, Phys. Lett. 109 A, 139
(1985).




\bibitem{32} M. Giffon and E. Predazzi,\textit{Quantum mechanics and high
energy physics}, Lett. Nuovo Cimento \textbf{37}, 430 (1983).

\bibitem{33} A. Kempf, G. Mangano, and R. B. Mann, \textit{Hilbert space
representation of the minimal length uncertainty relation}, Phys. Rev. D 52,
1108-1118 (1995)

\bibitem{34} L. N. Chang, D. Minic, N. Okamura, and T. Takeuchi, \textit{%
Exact solution of the harmonic oscillator in arbitrary dimensions with
minimal length uncertainty relations}, Phys. Rev. D 66, 026003 (2002).

\bibitem{35} S. Benczik, L. N. Chang, D. Minic and T.Takeuchi, \textit{The
hydrogen atom with minimal length}, arXiv:hep-th/0502222.

\bibitem{36} S. Benczik, L. N. Chang, D. Minic, N. Okamura, S. Rayyan, T.
Takeuchi, \textit{Classical implications of the minimal length uncertainty
relation}, arXiv:hep-th/0209119.

\bibitem{37} Y. Yamaguchi, \textit{Quantum mechanics and quantum field
theories in the Quantized Space. I}, Prog. Theor. Physics, Vol.111 No.4, 545
(2004).

\bibitem{38} Y. Yamaguchi, \textit{Quantum mechanics and quantum field
theories in the Quantized Space. II}, Prog. Theor. Physics, Vol.111 No.4,
545 (2005).

\bibitem{DVA} D.V. Ahluwalia, Phys Lett. A 275 (2000) 31-35

\bibitem{39} S. Girvin and R. Prange, \textit{The quantum Hall effect},
Springer New York (1987).

\bibitem{40} J. Bellisard, A. van Elst and H. Schultz-Valdes, \textit{The
noncommutative geometry and the quantum Hall effect}, (1993);
con-mat/93011005.

\bibitem{41} Y. Q. Li and B. Chen, \textit{Quantum theory for mesoscopic
electric circuits}, Phys. Rev. \textbf{B 53}, 4027 (1996).

\bibitem{42} Y. Q. Li, \textit{Commutation Relations in Mesoscopic Electric
Circuits}, 2000; con-mat/0009352.

\bibitem{43} J. C. Flores, \textit{Mesoscopic circuits with charge
discreteness: Quantum transmission lines}, Phys. Rev. \textbf{B 64}, 235309
(2001).

\bibitem{44} J. C. Flores and C. A. Utreras, \textit{Mesoscopic circuits
with charge discreteness: Quantum current magnification for mutual
inductances}, Phys. Rev. \textbf{B 66}, 153410 (2002).

\bibitem{45} J. Magueijo and L. Smolin, Phys. Rev. Lett. \textbf{88},
190403(2002) [arXiv:hep-th/0112090]

\bibitem{46} G. Amelino-Camelia, Int. J. Mod. Phys. D11, 35, 2002,
[arXiv:gr-qc/0012051], Ohys. Lett. B510, 255-263, 2001.

\bibitem{47} J. Kowalski-Glikman, Phys. Lett. A286, 391-394, 2001,
[arXiv:hep-th/0102098]; N.R. Bruno, G. Amelino-Camelia, J. Kowalski-Glikman,
Phys. Lett. B522, 133-138,2001.

\bibitem{48} D. Kimberly, J. Magueijo and J. Medeiros [arXiv:gr-qc/0303067]

\bibitem{49} Sijie Gao and Xiaoning Wu, \textit{Position space of Doubly
Special Relativity} [arXiv:gr-qc/0311009]

\bibitem{50} A. A. Deriglazov, Doubly special relativity in position space
starting from the conformal group, hep-th/0409232.

\bibitem{51} J. Moffat, Int. J. of Physics D \textbf{2}, 351 (1993); J.
Moffat, Foundations of Physics, \textbf{23}, 411 (1993)

\bibitem{52} A. Albrecht and J. Magueijo, Phys. Rev. D \textbf{59}, 043516
(1999)

\bibitem{53} Sung Ku Kim, Sun Myong Kim, Chaiho Rim and Jae Hyung Yee,
\textit{Propagation of Light in Doubly Special Relativity}
[arXiv:gr-qc/0401078]

\bibitem{KGV} D. Grumiller, W. Kummer, D.V. Vassilevich, Ukr. J.
Phys. 48 (2003) 329-333;[arXiv:hep-th/0301061].


\bibitem{DVA2} D. V. Ahluwalia-Khalilova, Class. Quant. Grav. 22
(2005) 1433-1450;[arXiv:hep-th/050314].

\bibitem{RW} R. Schutzhold, W.G. Unruh, JETP Lett. 78: 431-435, 2003, Pisma Zh. Eksp. Teor. Fiz. 78: 899-903, 2003.

\bibitem{Chryss}C. Chryssomalakos, E. Okon, Int. J. Mod. Phys. D13: 2003-2034, 2004 ;[arXiv:hep-th/0410212].

\bibitem{CP} John Collins et al., Phys Rev. Lett. 93: 191301, 2004;[arXiv:gr-qc/0403053].

\bibitem{54} C.~Leiva and M.~S.~Plyushchay, \textit{Conformal Symmetry of
Relativistic and Nonrelativistic Systems and AdS/CFT Correspondence}, Annals
of Physics 307 0310 (2003) 372-391. [arXiv:hep-th/0301244].

\bibitem{55} C. Leiva, \textit{Conformal Generators and Doubly Special
Relativity Theories}, Mod. Phys. Lett. A, Vol. 20, No. 11 (2005) pp.
861-867, [arXiv: gr-qg/0405149].

\bibitem{56} C. Leiva \textit{(}{Extended Conformal Group and DSR Velocities
on the Physical Surface }, [arXiv: hep-th/0410159].

\bibitem{57} V. Fock, \textit{The theory of space-time and gravitation},
Pergamon Press,1964.

\bibitem{58} S.N. Manida, [arXiv:gr-qc/9905046]

\bibitem{59} Juan M. Romero and Adolfo Zamora, \textit{Snyder noncommutative
space-time from two-time physics}, [arXiv:hep-th/0408193]

\bibitem{Hor}P.A. Horvathy, Exotic Galilean symmetry and non-commutative
mechanics in mathematical and in condensed matter
physics.[arXiv:hep-th/0602133 ]


\bibitem{60} Stefan Groot Nibbelink and Maxim Pospelov \textit{Lorentz
Violation in Supersymmetric Field Theories}, [arXiv:hep-ph/0404271].

\bibitem{61} J. Gamboa, M. Loewe and F. M\'{e}ndez, \textit{Relativistic
Invariance, Multiverse and Quantum Field Theory}, [aXiv:hep-th/0311014]

\bibitem{62} J. M. Carmona, J.L. Cort\'{e}s, J. Gamboa and F. M\'{e}ndez,
\textit{Quantum theory of noncommutative fields}, JHEP 0303:058,2003 [arXiv:
hep-th/0301248]

\bibitem{63} D. Kimberly, J. Magueijo, J. Medeiros, \textit{Non-Linear
relativity in Position Space}, Phys. Rev. D 70 (2004) 084007,
[arXiv:gr-qc/0303067]



\end{thebibliography}
\end{document}